\documentclass[letterpaper, 10 pt, conference]{ieeeconf}  

\IEEEoverridecommandlockouts                              

\usepackage{cite}
\usepackage[scr=boondoxo,scrscaled=1.05]{mathalfa}
\usepackage{bm}
\usepackage{algorithm}
\usepackage{amsmath,amssymb,amsfonts}
\usepackage{url}
\usepackage{algorithmic}
\usepackage{graphicx}
\usepackage{textcomp}
\usepackage{xcolor}

\newtheorem{prop}{Proposition}
\newtheorem{remark}{Remark}

\newcommand{\eor}{\ensuremath{\hfill\blacklozenge}}

\DeclareMathOperator{\diag}{diag}
\usepackage{flushend}

\title{\LARGE \bf
A comparison between joint and dual UKF implementations for state estimation and leak localization in water distribution networks*
}

\author{Luis Romero-Ben$^{1}$, Paul Irofti$^{2}$, Florin Stoican$^{3}$ and Vicenç Puig$^{4}$
\thanks{*The authors thank the support received by the project “Romanian Hub for Artificial Intelligence - HRIA”, Smart Growth, Digitization and Financial Instruments Program, 2021-2027, MySMIS no. 351416; and the projects “Sustainable and learning-based management of large-scale multi-resource systems” - SEAMLESS, ref. PID2023-148840OB-I00, and “Sensor fusion for real-time monitoring of leaks in water distribution networks” - FUSAGUA, no. 202550E039, both funded by MCIU/ AEI /10.13039/501100011033/FEDER, UE.}
\thanks{$^{1}$Luis Romero-Ben is with the Department of Systems Eng., Automation and Industrial Informatics (ESAII),
        Universitat Politècnica de Catalunya - BarcelonaTech (UPC)
        , 08028 Barcelona, Spain
        {\tt\small luis.romero.ben@upc.edu}}%
\thanks{$^{2}$Paul Irofti is with LOS-CS-FMI,
University of Bucharest, Romania {\tt\small paul@irofti.net}} 
\thanks{$^{3}$Florin Stoican is with the  Deptartment of Automation Control and Systems Engineering, Politehnica University of Bucharest, 060042 Romania, 
{\tt\small florin.stoican@upb.ro}}
\thanks{$^{4}$Vicenç Puig is with the Institut de Robòtica i Informàtica Industrial, CSIC-UPC
, Barcelona, 08028 Spain, and with the Supervision, Safety and Automatic Control Research Center (CS2AC), Universitat Politècnica de Catalunya
, Terrassa, Barcelona, 08222 Spain, {\tt\small vicenc.puig@upc.edu}}
}

\begin{document}

\maketitle
\thispagestyle{empty}
\pagestyle{empty}

\begin{abstract}

The sustainability of modern cities highly depends on efficient water distribution management, including effective pressure control and leak detection and localization. Accurate information about the network hydraulic state is therefore essential. This article presents a comparison between two data-driven state estimation methods based on the Unscented Kalman Filter (UKF), fusing pressure, demand and flow data for head and flow estimation. One approach uses a joint state vector with a single estimator, while the other uses a dual-estimator scheme. We analyse their main characteristics, discussing differences, advantages and limitations, and compare them theoretically in terms of accuracy and complexity. Finally, we show several estimation results for the L-TOWN benchmark, allowing to discuss their properties in a real implementation.

\end{abstract}

\section{INTRODUCTION}

Clean water is crucial for the development of society, and water utilities seek to optimize its distribution. A major challenge degrading this delivery is leakage in water distribution networks (WDN), which is estimated to produce worldwide 126 billion cubic meters of lost water annually \cite{Liemberger2019}. Water utilities monitor leaks using both hardware-based and software-based methods. The former include night flow analysis for detection and acoustic sensing by human operators for localization. Software-based methods complement them, allowing to monitor wider areas through algorithms which analyse steady-state data from distributed sensors.

Software-based methods can be classified by their reliance on a hydraulic model, used to simulate potential leak scenarios. Model-based methods require well-calibrated models to compare simulated and measured data to localize leaks \cite{Perez2011}\cite{Steffelbauer2022}. 
Mixed model-based/data-driven methods reduce this reliance by only using the model to generate training data for learning-based algorithms \cite{Zhang2016,Irofti2023}. Purely data-driven schemes remove the model completely, relying instead on sensor data and structural information. They often use interpolation methods to estimate the full hydraulic state of the network under leak and leak-free conditions, which are then compared to identify the leak location \cite{Soldevila2020,RomeroBen2022}.

Most methods use nodal pressures or hydraulic heads (pressure + elevation) as state representatives, since pressure sensors have been historically cheaper and easier to install than others, and pressure-to-leak sensitivity is well-studied. Recently, demand meters (typically denoted as Automated Metering Reading - AMR) have become more common in WDNs due to the falling associated costs, motivating the need of integration of new data sources. To address this, sensor fusion methods have been recently used to improve state estimation in WNDs, such as non-linear Kalman filters like the Extended Kalman Filter (EKF) \cite{Jung2015} or the Unscented Kalman Filter (UKF) \cite{RomeroBen2024b}; and others like Factor Graph Optimization (FGO) \cite{Irofti2025}. In \cite{RomeroBen2025}, we presented an upgraded dual version of \cite{RomeroBen2024b} that estimates both heads and flows separately, using an UKF and a Kalman Filter respectively. However, the option of using a joint version of the UKF, as explained in \cite{VanDerMerwe2004}, was not explored.

This article compares the joint and dual formulations, outlining their conceptual differences along with theoretical strengths and limitations. They are defined in the context of state estimation in WDNs and evaluated over a well-known, realistic benchmark, i.e., L-TOWN \cite{Vrachimis2022}.

\noindent\textbf{Notation}: Scalars are represented with plain lowercase letters, while vectors and matrices are denoted in bold, through lowercase and uppercase letters, respectively. On the one hand, the $i$-th column of a matrix $\bm{A}$ is represented as $\bm{a}_i$, with $a_{ij}$ denoting the $i$-$j$ component. On the other hand, sub-indices $h$ and $q$ denote the link of a variable to head or flow respectively. Besides, $\bm{\hat{A}}$ is an approximation of $\bm{A}$. Note that $\left(\bm{x}\right)^{y}$  indicates that each element of $\bm{x}$ is raised to the power of $y$. For clarification, $n_s,\:n_q$ and $n_a$ are the number of pressure, flow and demand sensors, respectively.

\section{Preliminaries}

Sensor fusion methods integrate multiple sources of information to refine state estimation. Among them, Kalman-based filters are well-known for their efficacy across different fields. 
These filters are based on a recursive estimation process, using the previous estimated state and current measured data to update the state at the current iteration.

\subsection{Kalman Filter}\label{subsubsection:KF}

The linear Kalman Filter (KF) \cite{Kalman1960} computes optimal states through prediction and measurement update steps, assuming linearity and white Gaussian noise in both processes. Let us consider a linear system such as:

\begin{align}\label{eq:KF_preliminaries_linearsystem}
    \bm{x}^{[k]} &= \bm{F}\bm{x}^{[k-1]} + \bm{B}^{u}\bm{u}^{[k-1]} + \bm{w}^{[k-1]}, \\
    \bm{z}^{[k]} &= \bm{G}\bm{x}^{[k]} + \bm{v}^{[k]},
\end{align}

\noindent where $\bm{x}^{[k]}$ is the system's state at time instant $k$, $\bm{F}$ is the state transition matrix, $\bm{B}^{u}$ denotes the control-input matrix, $\bm{u}^{[k-1]}$ represents the control-input vector, $\bm{z}^{[k-1]}$ denotes the measurement vector and $\bm{G}$ is the measurement matrix. The process and measurement noises are $\bm{w}^{[k-1]} \sim \mathcal{N}(\bm{0},\bm{Q})$ and $\bm{v}^{[k]}\sim \mathcal{N}(\bm{0},\bm{R})$, where $\bm{Q}$ and $\bm{R}$ are the process and measurement noise covariance matrices, respectively. A step of the Kalman Filter procedure can be presented in Algorithm \ref{alg:EKF-prelimiaries} if $\bm{F}^{[k-1]}, \bm{G}^{[k]}$ are fixed, and taking 
\begin{align}
    \label{eq:fF}\mbox{\textbf{f}}(\bm{x}^{[k]},\bm{u}^{[k]})&=\bm{F}\bm{\hat{x}}^{[k-1]} + \bm{B}^u\bm{u}^{[k]}, \\
    \label{eq:gG}\mbox{\textbf{g}}(\bm{x}^{[k]})&=\bm{G}\bm{x}^{[k]}.
\end{align}


Note that $\bm{\hat{x}}^{[k]} = \mathbb{E}\left[\bm{x}^{[k]}\right]$ denotes the state estimation expected value and $\bm{\hat{P}}^{[k]}$ is the state error covariance matrix, computed as $\bm{\hat{P}}^{[k]} = \mathbb{E}\left[\left(\bm{x}^{[k]} - \bm{\hat{x}}^{[k]}\right)\left(\bm{x}^{[k]} - \bm{\hat{x}}^{[k]}\right)^{\top}\right]$. Besides, $\bm{K}^{[k]}$ is denoted as the Kalman gain.

\subsection{Extended Kalman Filter}

In \cite{Kalman1961}, the KF was extended to handle non-linear systems, which do not fulfill the optimality assumptions exposed in Section \ref{subsubsection:KF}. Let us consider a non-linear system as:
\begin{align}\label{eq:EKF_prelimiaries_nonlinear}
     \bm{x}^{[k]} &= \mbox{\textbf{f}}(\bm{x}^{[k-1]},\bm{u}^{[k-1]}) + \bm{w}^{[k-1]}, \\
     \bm{z}^{[k]} &= \mbox{\textbf{g}}(\bm{x}^{[k]}) + \bm{v}^{[k]},
\end{align}
\noindent where $\mbox{\textbf{f}}$ and $\mbox{\textbf{g}}$ denote the non-linear process and measurement functions respectively. The Extended Kalman Filter (EKF) linearizes the process and measurement models, computing the associated Jacobians as:  

\begin{equation}\label{eq:EKF_prelimiaries_Jacobian}
     \bm{F}^{[k-1]} = \left.\frac{\partial \mbox{\textbf{f}}}{\partial \bm{x}}\right|_{\bm{x}^{[k-1]},\bm{u}^{[k-1]}}, \;\;\;\;\;
     \bm{G}^{[k]} = \left.\frac{\partial \mbox{\textbf{g}}}{\partial \bm{x}}\right|_{\bm{x}^{[k]}_-}.
\end{equation}
\noindent A step of the EKF process is showed in Algorithm \ref{alg:EKF-prelimiaries}.

\begin{algorithm}[!ht]
\caption{Extended Kalman Filter}
\label{alg:EKF-prelimiaries}
\begin{algorithmic}[1]
\REQUIRE {$\bm{\hat{x}}^{[k-1]}, \bm{\hat{P}}^{[k-1]}, \bm{u}^{[k-1]}, \bm{z}^{[k]}, \bm{F}^{[k-1]}, \bm{G}^{[k]}, \bm{Q}, \bm{R}$}
\STATE $\bm{\hat{x}}^{[k]}_{-} = \mbox{\textbf{f}}(\bm{\hat{x}}^{[k-1]},\bm{u}^{[k-1]})$
\STATE $\bm{\hat{P}}^{[k]}_{-} = \bm{F}^{[k-1]}\bm{\hat{P}}^{[k-1]}(\bm{F}^{[k-1]})^{\top} + \bm{Q}$
\STATE $\bm{K}^{[k]} = \bm{\hat{P}}^{[k]}_{-}(\bm{G}^{[k]})^{\top}\left(\bm{G}^{[k]}\bm{\hat{P}}^{[k]}_{-}(\bm{G}^{[k]})^{\top} + \bm{R}\right)^{-1}$
\STATE $\bm{\hat{x}}^{[k]} = \bm{\hat{x}}^{[k]}_{-} + \bm{K}^{[k]}\left(\bm{z}^{[k]} - \mbox{\textbf{g}}\bigl(\bm{\hat{x}}^{[k]}_{-}\bigr)\right)$
\STATE $\bm{\hat{P}}^{[k]} = \left(\bm{I}_n - \bm{K}^{[k]}\bm{G}^{[k]}\right)\bm{\hat{P}}^{[k]}_{-}$
\RETURN $\bm{\hat{x}}^{[k]}, \bm{\hat{P}}^{[k]}$
\end{algorithmic}
\end{algorithm}

\subsection{Unscented Kalman Filter}

Despite its widespread use, the EKF's precision is limited by the use of first-order linearization, neglecting higher-order terms potentially leading to a divergence from the actual system behaviour. The UKF \cite{Julier1997} overcomes this limitation by avoiding linearization, without an extra computational cost. While the EKF approximates the non-linear models, the UKF approximates the distribution of the state, regarded as a random variable, propagating a set of deterministically selected points from the original distribution through the actual non-linear models, to then recover the mean and covariance by weighted averaging. These points, denoted as sigma points, are computed as follows:

\begin{equation}\label{eq:UKF_sigma_points}
    \bm{\mathcal{X}}^{[k]} =\begin{bmatrix}
    \bm{\hat{x}}^{[k]} & \bm{\hat{X}}^{[k]} + \eta\sqrt{\bm{\hat{P}}^{[k]}} & \bm{\hat{X}}^{[k]} - \eta\sqrt{\bm{\hat{P}}^{[k]}}
\end{bmatrix},
\end{equation}

\noindent where $\bm{\hat{X}}^{[k]}=\bm{\hat{x}}^{[k]}\bm{1}_{1\times n}$, with $\bm{1}_{1\times n}$ being a matrix of all ones of size $1\times n$; and $\eta=\sqrt{n+\lambda}$ and $\lambda = n(\alpha^2-1)$ are scaling parameters, with $n$ being the system's state size ($\alpha$ defines the spread of the sigma points around the mean).

If we rewrite \eqref{eq:UKF_sigma_points} as $\bm{\mathcal{X}}^{[k]} =\mbox{\textbf{SP}}(\bm{\hat{x}}^{[k]},\bm{\hat{P}}^{[k]},\eta)$, a step of the UKF procedure can be defined as presented by Algorithm \ref{alg:UKF-prelimiaries}.

\begin{algorithm}[!ht]
\caption{Unscented Kalman Filter}
\label{alg:UKF-prelimiaries}
\begin{algorithmic}[1]
\REQUIRE {$\bm{\hat{x}}^{[k-1]}, \bm{\hat{P}}^{[k-1]}, \bm{u}^{[k-1]}, \bm{z}^{[k]}, \bm{w}^{(m)}, \bm{w}^{(c)}, \eta, \bm{Q}, \bm{R}$}
\STATE $\bm{\mathcal{X}}^{[k-1]} = \mbox{\textbf{SP}}(\bm{\hat{x}}^{[k-1]},\bm{\hat{P}}^{[k-1]},\eta)$
\STATE $\bm{\mathcal{X}}^{[k]}_{-} = \mbox{\textbf{f}}(\bm{\mathcal{X}}^{[k-1]},\bm{u}^{[k-1]})$
\STATE $\bm{\hat{x}}^{[k]}_{-} = \sum_{i=0}^{2n} w_i^{(m)}\bm{\mathcal{X}}^{[k]}_{i,-}$
\STATE $\bm{\hat{P}}^{[k]}_{-} = \sum_{i=0}^{2n} w_i^{(c)}\left(\bm{\mathcal{X}}^{[k]}_{i,-} - \bm{\hat{x}}^{[k]}_{-}\right)\left(\bm{\mathcal{X}}^{[k]}_{i,-} - \bm{\hat{x}}^{[k]}_{-}\right)^{\top} + \bm{Q}$
\STATE $\bm{\mathcal{X}}^{[k]}_{-} = \mbox{\textbf{SP}}(\bm{\hat{x}}^{[k]}_{-},\bm{\hat{P}}^{[k]}_{-},\eta)$
\STATE $\bm{\mathcal{Y}}^{[k]}_{-} = \mbox{\textbf{g}}(\bm{\mathcal{X}}^{[k]}_{-})$
\STATE $\bm{\hat{y}}^{[k]}_{-} = \sum_{i=0}^{2n} w_i^{(m)}\bm{\mathcal{Y}}^{[k]}_{i,-}$
\STATE $\bm{\hat{P}}^{[k]}_{yy} = \sum_{i=0}^{2n} w_i^{(c)}\left(\bm{\mathcal{Y}}^{[k]}_{i,-} - \bm{\hat{y}}^{[k]}_{-}\right)\left(\bm{\mathcal{Y}}^{[k]}_{i,-} - \bm{\hat{y}}^{[k]}_{-}\right)^{\top} + \bm{R}$
\STATE $\bm{\hat{P}}^{[k]}_{xy} = \sum_{i=0}^{2n} w_i^{(c)}\left(\bm{\mathcal{X}}^{[k]}_{i,-} - \bm{\hat{x}}^{[k]}_{-}\right)\left(\bm{\mathcal{Y}}^{[k]}_{i,-} - \bm{\hat{y}}^{[k]}_{-}\right)^{\top}$
\STATE $\bm{K}^{[k]} = \bm{\hat{P}}^{[k]}_{xy}(\bm{\hat{P}}^{[k]}_{yy})^{-1}$
\STATE $\bm{\hat{x}}^{[k]} = \bm{\hat{x}}^{[k]}_{-} + \bm{K}^{[k]}\left(\bm{z}^{[k]} - \bm{\hat{y}}^{[k]}_{-}\right)$
\STATE $\bm{\hat{P}}^{[k]} = \bm{\hat{P}}^{[k]}_{-} - \bm{K}^{[k]}\bm{\hat{P}}^{[k]}_{yy}(\bm{K}^{[k]})^{\top}$
\RETURN $\bm{\hat{x}}^{[k]}, \bm{\hat{P}}^{[k]}$
\end{algorithmic}
\end{algorithm}

The weighted averaging (steps 3--4 and 7--9) is performed through $\bm{w}^{(m)}$ for mean-related computations and $\bm{w}^{(c)}$ for covariance-related steps, which are defined as $w_0^{(m)} = \frac{\lambda}{n+\lambda}, w_0^{(c)} = \frac{\lambda}{n+\lambda} + (1 - \alpha^2 + \beta^2)$ and $w_i^{(m)} = w_i^{(c)} = \frac{1}{2(n+\lambda)},\: \forall i=1,2,\ldots,2n$. Parameter $\beta$ can be used to add knowledge about the actual distribution of the state~\cite{Julier1997}.

\section{Methodology}

Recently, we applied the UKF to address state estimation in WDNs, designing an approach capable of integrating structural and physics-informed interpolation with pressure and demand data, referred to as UKF-AW-GSI \cite{RomeroBen2024b}, which was then extended to estimate both head and flow states in a dual manner, leading to Dual UKF-AW-GSI or D-UKF-AW-GSI \cite{RomeroBen2025}. Here, we present these implementations and a potential joint state version for head-flow estimation, which is then conceptually compared to the dual version. To this end, let us model the network structure through a simple, weighted and directed graph $\mathcal{G}=(\mathcal{V},\mathcal{E})$.  $\mathcal{V}$ denotes the set of nodes, including reservoirs and junctions of the WDN, where $n_{\mathcal{V}} = |\mathcal{V}|$. $\mathcal{E}$ is the set of edges, i.e., WDN pipes, with $n_{\mathcal{E}} = |\mathcal{E}|$. The $i$-$th$ node is denoted as $\mathscr{v}_i\in\mathcal{V}$, and the $k$-$th$ edge is represented as $\mathscr{e}_k = \mathscr{e}_{ij} = (\mathscr{v}_i,\mathscr{v}_j)\in\mathcal{E}$, where $\mathscr{v}_i$ is the source node and $\mathscr{v}_j$ is the sink. 

\subsection{UKF-AW-GSI}

This method is designed to iteratively improve an initial state estimation guess by means of pressure and demand sensor fusion and a linear structural/physics-informed diffusion prediction process. This initial guess comes from an interpolation strategy denoted as Analytical Weighting Graph-based State Interpolation or AW-GSI \cite{Irofti2023}, which allows to start the estimation process from an already accurate state. 
This technique is based on the original Graph-based State Interpolation (GSI) method \cite{RomeroBen2022}, which estimates the WDN state (given by the nodal hydraulic heads) by means of the next optimization problem:

\begin{align}\label{eq:GSI_opt}
\min_{\bm{\hat{h}}, \gamma} \quad & \frac{1}{2}\big[\bm{\hat{h}}^T\bm{L}\bm{D}^{-2}\bm{L}\bm{\hat{h}}+\zeta \gamma ^2\big],\\
\label{eq:GSI_opt_b}\textrm{s.t.} \quad & \bm{M} \bm{\hat{h}}\leq \gamma \cdot\bm{1}_{n},\ \gamma > 0,\ \bm{S}\bm{\hat{h}}=\bm{\hat{h}}^{s},  
\end{align}

\noindent where $\bm{\hat{h}}\in \mathbb{R}^{n_{\mathcal{V}}}$ is the estimated state vector. 
$\bm L=\bm D - \bm W$ is the Laplacian of $\mathcal G$, with $\bm{W}$, $\bm{D}\in \mathbb{R}^{n_{\mathcal{V}} \times n_{\mathcal{V}}}$ being the weighted adjacency and degree matrices of $\mathcal G$, respectively. $\bm{M}\in \mathbb{R}^{n_{\mathcal{E}}\times n_{\mathcal{V}}}$ is the incidence matrix, $\bm S \in \mathbb{R}^{n_{s}\times n_{\mathcal{V}}}$ is the sensorization matrix (with $n_s$ being the number of pressure sensors) and $\bm{\hat{h}}^{s}\in \mathbb{R}^{n_s}$ is the vector of head measurements. The cost function in \eqref{eq:GSI_opt} pursues graph structural diffusion with the first sub-goal and directionality fitting with the second through the minimization of slack variable $\gamma$, complemented with the constraints in \eqref{eq:GSI_opt_b}.

The structural-based weights in $\bm W$ are upgraded through AW-GSI in \cite{Irofti2023}, leading to physics-informed weights that are derived from an explicit approximation of the implicit relation between neighbouring heads:

\begin{equation}\label{eq:AW_weights}
    w^{AW}_{ij}(\overline{h}_i,\overline{h}_j) = \tau_{kk}^{-0.54}\left[b_{kj}(\overline{h}_i-\overline{h}_j)\right]^{1-\frac{1}{1.852}}, 
\end{equation}

\noindent where $w^{AW}_{ij}$ is the $i$-$j$ element of $\bm{W}^{AW}$ (the weighted adjacency matrix for AW-GSI), $\overline{\bm h}$ is the leak-free estimated state vector, and $\tau_{kk}=(10.67 \rho_{k})/(\mu_{k}^{1.852} \delta_{k}^{4.87})$ is $k$-$th$ diagonal element of the resistance coefficient matrix $\bm{T}\in\mathbb{R}^{n_{\mathcal{E}}\times n_{\mathcal{E}}}$, in S.I. units. In this case, $b_{kj}$ is the $k$-$j$ element of a pressure-based incidence matrix $\bm B$, defined as: 

\begin{equation}\label{eq:incidence}
    b_{kj}=\begin{cases}-1,& h_i\geq  h_j\;\; (\mathscr{e}_k = (\mathscr{v}_i,\mathscr{v}_j)\in \mathcal{E}), \\
    \hphantom{-}1,& h_i < h_j \;\; (\mathscr{e}_k = (\mathscr{v}_j,\mathscr{v}_i)\in \mathcal{E}),\\ 
    \hphantom{-}0,& (\mathscr{v}_i,\mathscr{v}_j)\notin \mathcal{E}\; \mbox{and} \;(\mathscr{v}_j,\mathscr{v}_i)\notin \mathcal{E}.\end{cases}
\end{equation}

The physics-informed weights in $\bm{W}^{AW}$ are used within AW-GSI to retrieve the initial guess for UKF-AW-GSI. The UKF process requires the definition of the prediction and data assimilation functions. The process function of the prediction stage is designed to perform graph diffusion of nodal information to neighbouring nodes, iteration by iteration, whereas the data assimilation step is defined through a non-linear function relating demands and heads. UKF-AW-GSI can be represented by Algorithm \ref{alg:J-UKF(AW)GSI}\footnote{Note that processes in steps 3, 4 and 5 are represented by the notation $[\cdot,\cdot,\ldots]\leftarrow(\cdot,\cdot,\ldots)$ for simplicity, where inputs and outputs are indicated.} if all the $\bm{\hat{x}}$ related variables are only linked to $\bm{\hat{h}}$, and removing steps 8-11. The method has three main stages coming from Algorithm \ref{alg:EKF-prelimiaries} (with \eqref{eq:fF} for the linear KF) and Algorithm \ref{alg:UKF-prelimiaries}:


\subsubsection{State prediction} UKF-AW-GSI implements graph diffusion through a linear process function, i.e., $\bm{\hat{h}}^{[k]} = 
    \bm{F}_h\bm{\hat{h}}^{[k-1]}$, 
where $\bm F _h= \frac{n_a}{n_{\mathcal{V}}}\left(\bm I_{n_{\mathcal{V}}}- \bm{\Phi}^{-1}\bm{\Omega}\right)+\bm{\Phi}^{-1}\bm{\Omega}$, with $\bm{\Phi}=\bm{D}^{AW}$ and $\bm{\Omega}=\bm{W}^{AW}$ ($n_a$ is the number of AMRs). 

\begin{prop}\label{prop_1}
    The utilization of steps 1-4 of Algorithm \ref{alg:UKF-prelimiaries} or steps 1-2 of Algorithm \ref{alg:EKF-prelimiaries} (considering the linear KF in \eqref{eq:fF}) is equivalent if the process function is linear. 
\end{prop}
\begin{proof} See the Appendix.
\end{proof}

\subsubsection{Measurement propagation} the data assimilation function implements both the head measurement function and the demand-head relation as follows:

\begin{equation}\label{eq:UKFGSI_data_assimilation}
     \bm{y}^{[k]} = \mbox{\textbf{g}}_h(\bm{h}^{[k]}) = \begin{bmatrix}
         \bm{S}\bm{h}^{[k]} \\ -\left(\bm{B}^{[k]}_c\right)^{\top}\mkern-12mu\left(\bm{T}^{-1}\bm{B}^{[k]}\bm{h}^{[k]}\right)^{\frac{1}{1.852}}
     \end{bmatrix},
\end{equation}

\noindent where $\bm{B}^{[k]}$ is obtained from $\bm{h}^{[k]}$ using \eqref{eq:incidence}, $\bm{B}^{[k]}_c$ is the sub-matrix formed from its columns that correspond to the nodes with demand sensors. 

\subsubsection{Correction} during this process, the measurement vector is settled as $\bm z = [\bm{h}_s^{\top}\;\; \bm{c}_a^{\top}]^{\top}$, with $\bm{c}_a$ being the vector of demand measurements.


\subsection{D-UKF-AW-GSI}

This approach estimates both head and flow states, using two separated estimators in a dual manner \cite{RomeroBen2025}. To this end, UKF-AW-GSI is paired with a linear KF that estimates flows, linking them through ``virtual measurements", i.e., fixed values from the current estimation of the other estimator. The algorithm of D-UKF-AW-GSI is presented in Algorithm \ref{alg:D-UKF(AW)GSI}.

\begin{algorithm}[t]
\caption{D-UKF-AW-GSI}
\label{alg:D-UKF(AW)GSI}
\begin{algorithmic}[1]
\REQUIRE {$\bm{\hat{h}}^{[0]}, \bm{\hat{P}}_h^{[0]}, \bm{\hat{P}}_q^{[0]}, \bm{z}, \bm{q}_s, \bm{F}_q, \bm{G}_q, \bm F_h, \mbox{\textbf{g}}, \bm{w}^{(m)}, \bm{w}^{(c)}, \eta,$ \\ $\;\;\;\;\;\;\;\;\;\bm{Q}_h, \bm{R}_h,\bm{Q}_q, \bm{R}_q, k_{ex}$}
\STATE Initialize $\bm{\hat{q}}^{[0]}=\left(\bm{T}^{-1}\bm{B}^{[0]}\bm{\hat{h}}^{[0]}\right)^{\frac{1}{1.852}}$, $k=1$, $\delta_{c} = $ False
\STATE Set $\bm z_h = \left[\bm z^{\top}\;\;(\bm{\hat{q}}^{[0]})^{\top}\right]^{\top}$, $\bm z_q = \left[\bm q_s^{\top}\;\;(\bm{\hat{q}}^{[0]})^{\top}\right]^{\top}$
\WHILE{$\delta_{c} = $ False} 
\STATE \textbf{State prediction} (steps 1-2 Algorithm \ref{alg:EKF-prelimiaries} \eqref{eq:fF}): \\ 
$\left[\bm{\hat{h}}^{[k]}_{-},\bm{\hat{P}}^{[k]}_{h,-}\right] \leftarrow \left(\bm{\hat{h}}^{[k-1]},\bm{\hat{P}}^{[k-1]}_h, \bm F_h,\bm{Q}_h\right)$ \\
$\left[\bm{\hat{q}}^{[k]}_{-},\bm{\hat{P}}^{[k]}_{q,-}\right] \leftarrow \left(\bm{\hat{q}}^{[k-1]},\bm{\hat{P}}^{[k-1]}_q, \bm F_q, \bm{Q}_q\right)$ 
\STATE \textbf{Measurement propagation} (steps 5-9 of Algorithm~\ref{alg:UKF-prelimiaries}): \\ 
\small$\left[\bm{\hat{y}}^{[k]}_{-},\bm{\hat{P}}^{[k]}_{yy}, \bm{\hat{P}}^{[k]}_{xy}\right] \leftarrow \left(\bm{\hat{h}}^{[k]}_{-},\bm{\hat{P}}^{[k]}_{-},\mbox{\textbf{g}}, \bm{R}, \bm{w}^{(m)}, \bm{w}^{(c)}, \eta\right)$  
 \STATE \normalsize\textbf{Correction} (steps 10-12 Alg.~\ref{alg:UKF-prelimiaries} / steps 3-5 Alg.~\ref{alg:EKF-prelimiaries} \eqref{eq:gG}):\\ 
$\left[\bm{\hat{h}}^{[k]}, \bm{\hat{P}}^{[k]}_h\right] \leftarrow \left(\bm{\hat{h}}^{[k]}_{-},\bm{\hat{P}}^{[k]}_{h,-}, \bm{\hat{P}}^{[k]}_{yy}, \bm{\hat{P}}^{[k]}_{xy}, \bm{z}_h\right)$ \\ 
$\left[\bm{\hat{q}}^{[k]}, \bm{\hat{P}}^{[k]}_q\right] \leftarrow \left(\bm{\hat{q}}^{[k]}_{-},\bm{\hat{P}}^{[k]}_{q,-}, \bm{G}_q, \bm{R}_q, \bm{z}_q\right)$
\IF{$\mbox{mod}(k,k_{ex}) = 0$}
    \STATE   
    $\bm{z}_h = \begin{bmatrix} \bm z^{\top} & (\bm{\hat{q}}^{[k]})^{\top}\end{bmatrix}^{\top}$ \\
    $\bm{z}_q = \begin{bmatrix} \bm{q}_s^{\top} & \left(\left(\bm{T}^{-1}\bm{\hat{B}}^{[k]}\bm{\hat{h}}^{[k]}\right)^{\frac{1}{1.852}}\right)^{\top}\end{bmatrix}^{\top}$
\ENDIF
\STATE $\delta_{c} =$ convergence\_criteria($\bm{\hat{h}}^{[k]},\bm{\hat{h}}^{[k-1]}$)
\STATE $k = k + 1$
\ENDWHILE
\RETURN $\bm{h}^{[k]},\; \bm{q}^{[k]}$
\end{algorithmic}
\end{algorithm}

The flow state is initially retrieved from the head initial guess through Hazen-Williams. Then, the measurement vectors can be defined: (i) the (head) UKF measurement vector, containing the head and demand data in $\bm z$, and the initial flow guess; (ii) the (flow) KF measurement vector, containing flow readings and the initial flow guess. Then, both UKF and KF start iterating, updating the measurement vectors with a certain frequency $k_{ex}$. In this way, both estimators synchronize their operation, leading to a consistent head/flow solution. For the KF, the state propagation and measurement update matrices are $\bm F_q=\bm I_{n_{\mathcal{E}}}$ and $\bm{G}_q = [\bm{S}_q^{\top} \;\;\;\bm{I}_{n_{\mathcal{E}}}^{\top}]^{\top}$.

\begin{remark}
    The convergence criteria mentioned in step 10 of Algorithm \ref{alg:D-UKF(AW)GSI} is user-defined. A simple and common option is to define a number of maximum iterations, although more complex criteria regarding the analysis of the state difference between iterations can be implemented.\eor
\end{remark}

\subsection{J-UKF-AW-GSI}

This method defines a composited or joint state, i.e.,  $\bm{x}^{[k]} = \begin{bmatrix}
        (\bm{h}^{[k]})^{\top} & (\bm{q}^{[k]})^{\top}
\end{bmatrix}^{\top}$, which is estimated through Algorithm \ref{alg:J-UKF(AW)GSI}. In this case, $\bm F_x, \bm Q_x$ and $\bm R_x$ are block diagonal matrices, with $\bm F_x=diag(\bm F_h,\bm F_q)$, $\bm Q_x=diag(\bm Q_h,\bm Q_q)$ and $\bm R_x=diag(\bm R_h,\bm R_q)$. Moreover, $\bm{z}_m = \begin{bmatrix} \bm{h}_s^\top &
\bm{q}_s^\top &
\bm{c}_a^\top\end{bmatrix}^\top$.
The measurement function $\mbox{\textbf{g}}_x$ is defined as\footnote{Function $\mbox{\textbf{g}}_x$ is expressed in terms of $\bm{\hat{h}}^{[k]}$ and $\bm{\hat{q}}^{[k]}$ for readability. All the sub-functions must be operating over $\bm{\hat{x}}^{[k]}$ in the actual implementation.}:

\begin{equation}\label{eq:J-UKFGSI_data_assimilation}
     \bm{y}^{[k]} = \mbox{\textbf{g}}_x(\bm{x}^{[k]}) = \begin{bmatrix}
         \bm{S}\bm{h}^{[k]} \\ \bm{S}_q\bm{q}^{[k]} \\ -\left(\bm{B}^{[k]}_c\right)^{\top}\mkern-12mu\left(\bm{T}^{-1}\bm{B}^{[k]}\bm{h}^{[k]}\right)^{\frac{1}{1.852}} \\
         \left(\bm{T}^{-1}\bm{B}^{[k]}\bm{h}^{[k]}\right)^{\frac{1}{1.852}}\\ \bm{q}^{[k]}
     \end{bmatrix},
\end{equation}

\noindent where $\bm{S}_q\in\mathbb{R}^{n_q\times n_{\mathcal{E}}}$ is the flow sensorization matrix, with $n_q$ being the number of flow sensors. The last two entries of $\mbox{\textbf{g}}_x$ implicitly impose a relation between estimated flow states and flow states  reconstructed from heads, which is parallel to the ``virtual measurements" scheme of the dual approach. 

\begin{algorithm}[t]
\caption{J-UKF-AW-GSI}
\label{alg:J-UKF(AW)GSI}
\begin{algorithmic}[1]
\REQUIRE {$\bm{\hat{x}}^{[0]}, \bm{\hat{P}}^{[0]}_x, \bm{z}_m, \bm F_x, \mbox{\textbf{g}}_x, \bm{w}_x^{(m)}, \bm{w}_x^{(c)}, \eta, \bm{Q}_x, \bm{R}_x, k_{ex}$}
\STATE Extract $\left[\bm{\hat{h}}^{[0]},\bm{\hat{q}}^{[0]}\right]\leftarrow \bm{\hat{x}}^{[0]}$
\STATE Initialize $k=1$, $\delta_{c} = $ False
\STATE Set $\bm z_x = \left[\bm z^{\top}_m\;\;(\bm{\hat{q}}^{[0]})^{\top} \;\;(\bm{\hat{q}}^{[0]})^{\top}\right]^{\top}$
\WHILE{$\delta_{c} = $ False} 
\STATE \textbf{State prediction} (steps 1-2 of Algorithm~\ref{alg:EKF-prelimiaries} \eqref{eq:fF}): \\ 
$\left[\bm{\hat{x}}^{[k]}_{-},\bm{\hat{P}}^{[k]}_{x,-}\right] \leftarrow \left(\bm{\hat{x}}^{[k-1]},\bm{\hat{P}}^{[k-1]}_x, \bm F_x,\bm{Q}_x\right)$  
\STATE \textbf{Measurement propagation} (steps 5-9 of Algorithm~\ref{alg:UKF-prelimiaries}): \\ \small
$\left[\bm{\hat{y}}^{[k]}_{-},\bm{\hat{P}}^{[k]}_{yy}, \bm{\hat{P}}^{[k]}_{xy}\right] \leftarrow \left(\bm{\hat{x}}^{[k]}_{-},\bm{\hat{P}}^{[k]}_{x,-},\mbox{\textbf{g}}_x, \bm{R}_x, \bm{w}_x^{(m)}, \bm{w}_x^{(c)}, \eta\right)$ 
 \STATE \normalsize\textbf{Correction} (steps 10-12 of Algorithm~\ref{alg:UKF-prelimiaries}):\\ 
$\left[\bm{\hat{x}}^{[k]}, \bm{\hat{P}}_x^{[k]}\right] \leftarrow \left(\bm{\hat{x}}^{[k]}_{-},\bm{\hat{P}}^{[k]}_{x,-}, \bm{\hat{P}}^{[k]}_{yy}, \bm{\hat{P}}^{[k]}_{xy}, \bm{z}_x\right)$ 
\IF{$\mbox{mod}(k,k_{ex}) = 0$}
    \STATE Extract $\left[\bm{\hat{h}}^{[k]},\bm{\hat{q}}^{[k]}\right]\leftarrow \bm{\hat{x}}^{[k]}$
    \STATE $\bm z_x = \left[\bm z^{\top}_m\;\;(\bm{\hat{q}}^{[k]})^{\top} \;\;\left(\left(\bm{T}^{-1}\bm{\hat{B}}^{[k]}\bm{\hat{h}}^{[k]}\right)^{\frac{1}{1.852}}\right)^{\top}\right]^{\top}$
\ENDIF
\STATE $\delta_{c} =$ convergence\_criteria($\bm{\hat{x}}^{[k]},\bm{\hat{x}}^{[k-1]}$)
\STATE $k = k + 1$
\ENDWHILE
\RETURN $\bm{x}^{[k]}$
\end{algorithmic}
\end{algorithm} 


\subsection{Theoretical discussion}\label{subsection:theoretical_discussion}

We now analyze the key differences between the joint and dual approaches discussed in this work, beginning with their respective covariance modeling. In the joint formulation, the full state vector combines both the head and flow states, resulting in a covariance matrix that can be expressed as a block matrix with the following structure:

\begin{equation}\label{eq:covariance_joint}
    \bm{\hat{P}}^{[k]}_x = \mathbb{E}\left[\left(\bm{x}^{[k]} - \bm{\hat{x}}^{[k]}\right)\left(\bm{x}^{[k]} - \bm{\hat{x}}^{[k]}\right)^{\top}\right]=\begin{bmatrix}
        \bm{\hat{P}}^{[k]}_h & \bm{\hat{P}}^{[k]}_{hq} \\ \bm{\hat{P}}^{[k]}_{qh} & \bm{\hat{P}}^{[k]}_{q}
    \end{bmatrix},
\end{equation}

\noindent with $\bm{\hat{x}}^{[k]} = \mathbb{E}\left[\bm{x}^{[k]}\right]$. $\bm{\hat{P}}^{[k]}_h$ and $\bm{\hat{P}}^{[k]}_q$ are the error covariance matrices corresponding to the head and flow states respectively, where $\bm{\hat{P}}^{[k]}_{hq}$ and $\bm{\hat{P}}^{[k]}_{qh}$ are the cross-covariance terms relating the head and flow states, and viceversa. In the case of the dual approach, each estimator operates over a specific hydraulic variable, and therefore only the diagonal terms of the block matrix in \eqref{eq:covariance_joint} are computed.
Thus, on the one hand, the joint covariance matrix in \eqref{eq:covariance_joint} allows to compute and operate over both the individual covariance matrices of the two state vectors and the cross-covariance terms correlating them and modelling their interactions. On the other hand, the dual approach is decoupling the state estimation process, fixing the non-computed state in each estimator while estimating the other. This is equivalent to setting $\bm{\hat{P}}^{[k]}_{hq} = \bm{\hat{P}}^{[k]}_{qh} = \bm{0}_{n\times n}$. Consequently, the dual version misses information regarding the correlation between the hydraulic states during covariance update, which may affect the final estimation from a theoretical perspective, with a slower convergence or even a reduced robustness if the correlations are high. However, analysing \eqref{eq:J-UKFGSI_data_assimilation} and the definition of the process model in J-UKF-AW-GSI, we see how the interactions between the head and flow states are neglected within the prediction function (for the sake of the comparison with D-UKF-AW-GSI), linking them through the ``virtual measurements". Thus, the cross-covariance terms are negligible, and the estimation accuracy of both methods should be comparable.


From the point of view of computational complexity, let us study the different methods mentioned in this Section. Analysing Algorithms \ref{alg:D-UKF(AW)GSI} and \ref{alg:J-UKF(AW)GSI}, we realize how the complexity of these algorithms is shaped by the UKF-related operations, specifically matrix squared-root computation (used for the "sigma point" generation process \eqref{eq:UKF_sigma_points}) and matrix inversion (used to compute the Kalman gain)\footnote{Note that the matrix inversion is also required to compute the Kalman gain in the linear KF algorithm, and hence this operation is used for the flow KF in D-UKF-AW-GSI.}, both solved using algorithms such as Cholesky decomposition, requiring $\frac13 n^3$ instructions, where $n$ is the size of the considered matrix. In the case of the UKF, the matrix inversion is applied over $\hat{\bm{P}}_{yy}$, whereas the squared-root is applied over $\hat{\bm{P}}_{-}$. For the linear KF, the matrix inversion operation is performed over a matrix computed as $\bm{E}=\bm{G}^{[k]}\bm{\hat{P}}^{[k]}_{-}(\bm{G}^{[k]})^{\top} + \bm{R}$. Thus, the dominant operation in each specific methodology, and consequently, the actual complexity of the method, is defined by the relation between the network dimensions and the sensorization densities. 

On the one hand, D-UKF-AW-GSI is composed of the head UKF and the flow KF. The UKF complexity is governed by the matrix inversion of $\bm{\hat{P}}_{yy}^h\in\mathbb{R}^{n_{z_h}\times n_{z_h}}$, with $n_{z_h}=n_s+n_a+n_{\mathcal{E}}$, and the matrix square-root computed over $\bm{\hat{P}}_{h,-}\in\mathbb{R}^{n_{\mathcal{V}}\times n_{\mathcal{V}}}$. The flow KF complexity mostly depends on the inversion of $\bm{E}_q\in\mathbb{R}^{n_{z_q}\times n_{z_q}}$, with $n_{z_q}=n_q+n_{\mathcal{E}}$. Analyzing these matrices, we see how the computational cost of D-UKF-AW-GSI will be likely governed by the inversion of $\bm{\hat{P}}_{yy}^h$: due to graph theory, $n_{\mathcal{E}}\geq n_{\mathcal{V}}-1$, so in any case where at least $n_s,n_a,n_q \geq 1$, matrix inversion would cost more than matrix squared-root computation. Besides, in the leak localization context, $n_s>n_a, n_q$.

On the other hand, J-UKF-AW-GSI's complexity is governed by the inversion of $\bm{\hat{P}}_{yy}^x\in\mathbb{R}^{n_{z_x}\times n_{z_x}}$, with $n_{z_x}=n_s+n_q+n_a+2n_{\mathcal{E}}$, and the matrix square-root computed over $\bm{\hat{P}}_{x,-}\in\mathbb{R}^{n_{\mathcal{V}}+n_{\mathcal{E}}\times n_{\mathcal{V}}+n_{\mathcal{E}}}$. In real-world networks, $n_s, n_a << n_{\mathcal{V}}$ and $n_q << n_{\mathcal{E}}$ due to the cost of installing and maintaining sensors. Besides, to analyze the relation between $n_{\mathcal{V}}$ and $n_{\mathcal{E}}$, we must consider $n_{\mathcal{E}}\geq n_{\mathcal{V}}-1$ for $\mathcal{G}$ to be connected. Thus the operation's complexity is normally shaped by the matrix inversion, considering that $n_{z_x}^{min}=n_s+n_q+n_a+2\left(n_{\mathcal{V}}-1\right)$ is higher than $n_{\mathcal{V}}+n_{\mathcal{E}}$ for sufficiently-sensorized networks.

Finally, comparing matrix sizes and considering the $\mathcal{O}(n^3)$ complexity of the involved operations, we conclude that J-UKF-AW-GSI is more computationally complex than D-UKF-AW-GSI.

\section{Case study and discussion\protect\footnotemark}

\footnotetext{Code and data: https://github.com/luisromeroben/JvsD-UKF-AW-GSI}

\begin{figure}[thpb]
    \centering
    \includegraphics[width=\linewidth]{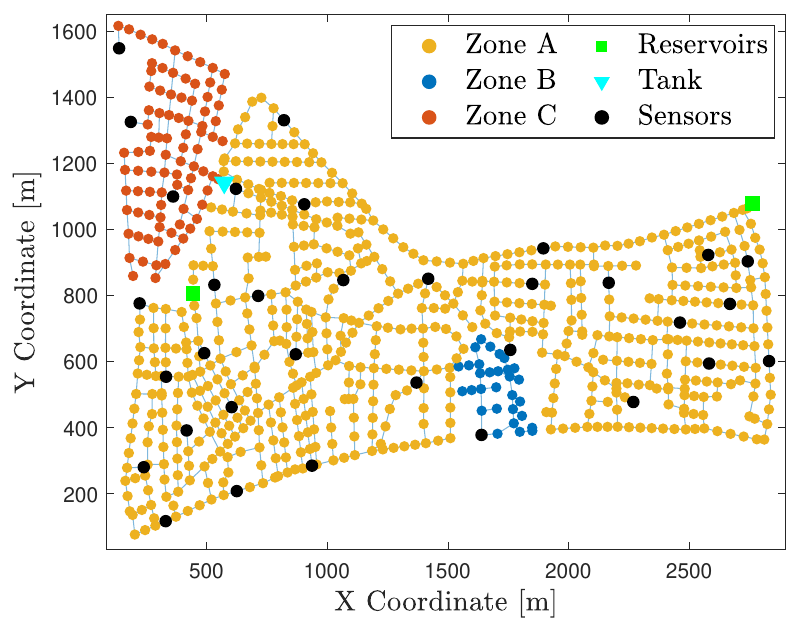}
    \caption{Scheme of the L-TOWN network.}
    \label{fig:network}
\end{figure}

With the aim of complementing the theoretical analysis, a performance comparison between D-UKF-AW-GSI and J-UKF-AW-GSI is carried out in a realistic benchmark. Specifically, we have selected the L-TOWN benchmark from the Battle of Leakage Detection and Isolation Methods 2020 - BattLeDIM2020 \cite{Vrachimis2022}. L-TOWN network is formed by 782 junctions, 905 pipes, 2 water inlets and 1 tank, with total pipe length of 42.6 km. A graphical representation of L-TOWN is depicted in Fig.~\ref{fig:network}. The network is split into three areas, depending on the nodal elevation:
\textit{Area A}, with 657 nodes (elevation between 16 and 48 meters, yellow in Fig.~\ref{fig:network}), connection to the 2 water inlets, 29 pressure sensors and 3 flow meters; \textit{Area B}, with 31 nodes (below 16 meters, blue in Fig.~\ref{fig:network}), 1 pressure meter and supplied with water from \textit{Area A} through a Pressure Reduction Valve (PRV); and
\textit{Area C}, with 94 nodes (above 48 meters, red in Fig.~\ref{fig:network}), 3 pressure sensors, 82 AMRs and supplied with water from \textit{Area A} through a tank.

To simplify the analysis, we focus on \textit{Area A}, as this zone is sufficiently large and equipped with pressure sensors (4.41\% of the nodes). Some adjustments are required to apply the UKF-based methods. First, the network graph is modified to eliminate the large head difference between the water inlets (100 m) and the rest of the WDN ($<75$ m), removing reservoir nodes and setting their neighbours as water inlets (with a height of 76 m), in order to avoid degraded estimation performance from the UKF-based methods. Besides, a set of 100 AMRs are supposed to be installed throughout \textit{Area A} to have access to demand measurements. They are placed using a model-free sensor placement method \cite{RomeroBen2022sp}. Besides, with the aim of reducing computational cost, we have considered a subset of 50 leak scenarios from the set of potential leak locations (excluding sensor locations)\footnote{To generate hydraulic data, the Dataset Generator from BattLeDIM2020 is used. Each leak scenario is simulated for 5 minutes to reduce computation time, with a leak diameter of 2 cm.}. The following values have been assigned to the parameters of the UKF-based methods, considering the degree of confidence in each one of the information sources: (i) for D-UKF-AW-GSI, $\bm{R}_h = \diag\left(10^{-4}\bm{I}_{n_s},10^{-4}\bm{I}_{n_a},10^{3}\bm{I}_{n_{\mathcal{E}}}\right)$, $\bm{R}_q = \diag\left(10^{-6}\bm{I}_{n_q},10^{-5}\bm{I}_{n_{\mathcal{E}}}\right)$, $\bm{Q}_h = \diag\left(\bm{I}_{n_{\mathcal{V}}}\right)$ and $\bm{Q}_q = \diag\left(10^{-5}\bm{I}_{n_{\mathcal{E}}}\right)$; (ii) for J-UKF-AW-GSI, $\bm{R}_x = \diag\left(10^{-4}\bm{I}_{n_s},10^{-6}\bm{I}_{n_q},10^{-4}\bm{I}_{n_a},10^{3}\bm{I}_{n_{\mathcal{E}}},10^{-5}\bm{I}_{n_{\mathcal{E}}}\right)$ and $\bm{Q}_x = \diag\left(\bm{I}_{n_{\mathcal{V}}},10^{-5}\bm{I}_{n_{\mathcal{E}}}\right)$; and (iii) common parameters such as $\alpha=10^{-3}$, $\beta=2$ (both standard in UKF implementations), and $k_{ex}=1$. Regarding the convergence criteria, a maximum number of iterations $k_{max}$ is defined to stop the UKF process.

Estimation results are presented through the root-squared-mean error
($RMSE(\bm x, \hat{\bm x}) = \sqrt{\frac{1}{n} \sum_{i=1}^{n} (x_i - \hat{x}_i)^2}$, where $\bm{x}$ is a vector) in terms of head ($RMSE_h$), and flow ($RMSE_q$) in Table \ref{table:estimation_results_jointvsdual}. Besides, computation time $t$ is shown as an indicator of computational complexity. These results cover a range of values of $k_{max}$, showing its effect on the accuracy and computation cost.

\begin{figure}[thpb]
    \centering
    \includegraphics[width=\linewidth]{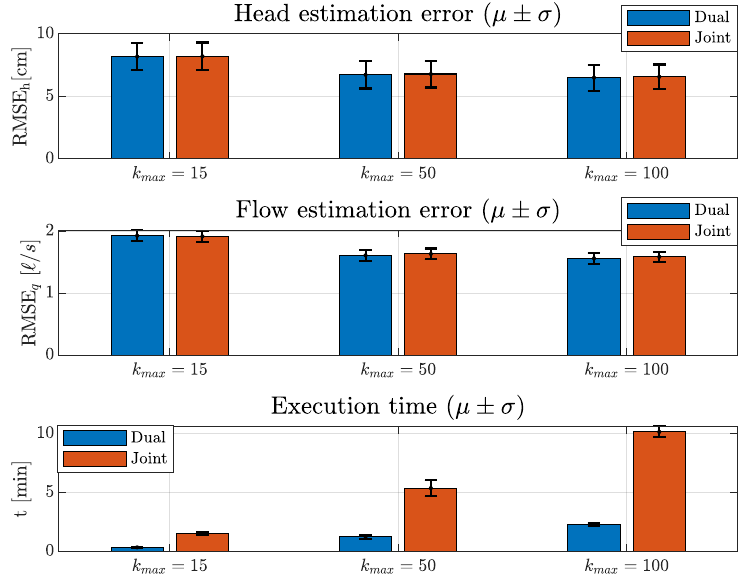}
    \caption{Estimation results for D-UKF-AW-GSI and J-UKF-AW-GSI in \textit{Area A}, in terms of head and flow estimation error, and execution time, for different values of $k_{max}$.}
    \label{fig:results}
\end{figure}

\begin{table}[h]
\caption{{Estimation performance comparison between D-UKF-AW-GSI and J-UKF-AW-GSI in \textit{Area A} of L-TOWN.}}
\label{table:estimation_results_jointvsdual}
\begin{center}
\begin{tabular}{|c|ccc|}
\hline
$k_{max}$ & $RMSE_h$ (cm) & $RMSE_q$ $(\ell/s)$ & $t$ (min)\\
\hline
\multicolumn{4}{|c|}{\textbf{J-UKF-AW-GSI}}\\
\hline
 15 & $8.20 \pm 1.11$ & $1.91 \pm 0.09$ & $1.51 \pm 0.10$\\
 50 & $6.79 \pm 1.08$ & $1.63 \pm 0.09$ & $5.37 \pm 0.68$\\
 100 &  $6.56 \pm 0.98$ & $1.59 \pm 0.08$ & $10.14 \pm 0.47$ \\
\hline
\multicolumn{4}{|c|}{\textbf{D-UKF-AW-GSI}}\\
\hline
 15 & $8.18 \pm 1.11$ & $1.93 \pm 0.09$ & $0.34 \pm 0.02$\\
 50 & $6.73 \pm 1.10$ & $1.61 \pm 0.09$ & $1.24 \pm 0.16$\\
 100 &  $6.50 \pm 1.04$ & $1.56 \pm 0.09$ & $2.29 \pm 0.14$ \\
\hline
\end{tabular}
\end{center}
\end{table}
The comparison reveals that both approaches achieve similar estimation accuracy, with the dual formulation generally performing slightly better in head estimation. However, the dual approach yields a substantial reduction in computation time—approximately 77\% on average. These results are illustrated in Fig.~\ref{fig:results}, where the bars represent the mean performance for each combination of method and $k_{\max}$ value, and the vertical error bars indicate the corresponding standard deviations.


Regarding execution time, the observed difference between the joint and dual forms is consistent with the theoretical analysis derived in Section \ref{subsection:theoretical_discussion}. Besides, the results show a slightly better estimation performance of D-UKF-AW-GSI in comparison to J-UKF-AW-GSI. 
Several practical factors may contribute to slightly degrade the joint method's performance. Firstly, the increased state dimension in the joint version can lead to a more complex error propagation (due to rounding or numerical stability issues). Secondly, different scales between the composited states, which are heads and flows in our case, may deform the generated sigma points, leading to a loss of precision in the estimation of the mean and covariance of the joint state \cite{Yu2016}.  


\section{Conclusions}

The article objective to perform an in-depth analysis of the joint and dual implementation of UKF-based methods for state estimation and sensor fusions in water distribution networks was reached. Even the J-UKF-AW-GSI generally uses more information in its state estimation, which should provide more accurate results, our implementation does not use an explicit relation of composited states, and therefore, the accuracy of the joint and dual versions should be analogue. In practice we have seen that D-UKF-AW-GSI converges faster than J-UKF-AW-GSI to an acceptable estimate (specifically, a 77\% average reduction in computation time is achieved with the dual version in our benchmark). If given more execution time J-UKF-AW-GSI will reach the same results, but we conclude that our benchmarks do not recommend it in practice. Besides, J-UKF-AW-GSI presents a slight degradation in accuracy with respect to D-UKF-AW-GSI ($\sim$0.7\% of RMSE reduction by using the dual version), which is consistent with implementation challenges reported in the literature Therefore, we expect for the results to be consistent across other tasks than state estimation such as leak localization.

Future work will include a wide analysis of this comparison in additional leak scenarios and benchmarks, as well as deep analysis of the effects of uncertainty and relative variable scaling within the results of both approaches. Moreover, we will study how to exploit the explicit state relations within the joint method, in order to use the cross-covariance terms and improve estimation accuracy.

The authors would like to thank reviewer 19 from \cite{RomeroBen2025} for his/her comments and sparking an interest in this direction.



\section*{APPENDIX}\label{sec:appendix}

\subsection{Proof of Proposition~\ref{prop_1}}

Let us analyse the case of a linear process function in the prediction step of the UKF, i.e.,  $\bm{x}^{[k+1]} = \mbox{\textbf{f}}(\bm{x}^{[k]})=\bm{F}\bm{x}^{[k]}$, where $\bm{u}^{[k]}=\bm{0}_n$ and $\bm{w}^{[k]}$ is neglected for simplicity. The state prediction stage of the UKF is as follows:

\begin{gather}
    \bm{\mathcal{X}}^{[k-1]} = \mbox{\textbf{SP}}(\bm{\hat{x}}^{[k-1]},\bm{\hat{P}}^{[k-1]},\eta) \\
    \bm{\mathcal{X}}^{[k]}_{-} = \mbox{\textbf{f}}(\bm{\mathcal{X}}^{[k-1]})=\bm{F}\bm{\mathcal{X}}^{[k-1]}\\ \label{eq:proof_1}
    \bm{\hat{x}}^{[k]}_{-} = \sum_{i=0}^{2n} w_i^{(m)}\bm{\mathcal{X}}^{[k]}_{i,-}=\sum_{i=0}^{2n} w_i^{(m)}\bm{F}\bm{\mathcal{X}}^{[k-1]}_{i}
\end{gather}

\begin{equation}\label{eq:proof_2}
\begin{split}
    \bm{\hat{P}}^{[k]}_{-} = \sum_{i=0}^{2n} w_i^{(c)}\left(\bm{\mathcal{X}}^{[k]}_{i,-} - \bm{\hat{x}}^{[k]}_{-}\right)\left(\bm{\mathcal{X}}^{[k]}_{i,-} - \bm{\hat{x}}^{[k]}_{-}\right)^{\top} + \bm{Q}=\\\sum_{i=0}^{2n} w_i^{(c)}\left(\bm{F}\bm{\mathcal{X}}^{[k-1]}_{i} - \bm{\hat{x}}^{[k]}_{-}\right)\left(\bm{F}\bm{\mathcal{X}}^{[k-1]}_{i} - \bm{\hat{x}}^{[k]}_{-}\right)^{\top} + \bm{Q}.
\end{split}
\end{equation}

First, analysing \eqref{eq:proof_1}, and considering that the Unscented Transform (UT) is defined so that $\bm{\hat{x}}^{[k]} = \sum_{i=0}^{2n} w_i^{(m)}\bm{\mathcal{X}}^{[k]}_{i}$ \cite{Julier1997}, we get that:

\begin{equation}\label{eq:proof_3}
    \bm{\hat{x}}^{[k]}_{-} = \sum_{i=0}^{2n} w_i^{(m)}\bm{F}\bm{\mathcal{X}}^{[k-1]}_{i}=\bm{F}\sum_{i=0}^{2n} w_i^{(m)}\bm{\mathcal{X}}^{[k-1]}_{i}=\bm{F}\bm{x}^{[k-1]},
\end{equation}

\noindent which is equivalent to the state prediction step of the linear Kalman  Filter. For the covariance update, analysing \eqref{eq:proof_2}, considering that the UT is defined so that $\bm{\hat{P}}^{[k]} = \sum_{i=0}^{2n} w_i^{(c)}\left(\bm{\mathcal{X}}^{[k]}_{i} - \bm{\hat{x}}^{[k]}\right)\left(\bm{\mathcal{X}}^{[k]}_{i} - \bm{\hat{x}}^{[k]}\right)^{\top} + \bm{Q}$, and inserting \eqref{eq:proof_3}, we get that:

\small
\begin{equation}
\begin{split}
    \bm{\hat{P}}^{[k]}_{-} = \sum_{i=0}^{2n} w_i^{(c)}\left(\bm{F}\bm{\mathcal{X}}^{[k-1]}_{i} - \bm{\hat{x}}^{[k]}_{-}\right)\left(\bm{F}\bm{\mathcal{X}}^{[k-1]}_{i} - \bm{\hat{x}}^{[k]}_{-}\right)^{\top} + \bm{Q} = \\ \bm{F}\sum_{i=0}^{2n} w_i^{(c)}\left(\bm{\mathcal{X}}^{[k-1]}_{i} - \bm{\hat{x}}^{[k-1]}\right)\left(\bm{\mathcal{X}}^{[k-1]}_{i} - \bm{\hat{x}}^{[k-1]}\right)^{\top}\bm{F}^{\top} + \bm{Q}=\\  \bm{F}\bm{\hat{P}}^{[k-1]}\bm{F}^{\top} + \bm{Q},
    \end{split}
\end{equation}
\normalsize

\noindent which is equivalent to the covariance prediction step of the linear Kalman Filter.

\bibliographystyle{IEEEtran}
\bibliography{references}

\end{document}